\documentclass[dvips]{article}

\usepackage{icrc1238}

\title{Test of the hadronic interaction models at around *10 TeV with Tibet EAS core data }

\newcommand{\etal}{\MakeLowercase{\textit{et al. }}} 
\newcommand{\e}{\MakeLowercase{\textit{e}}} 
\shorttitle{J.SHAO \etal test of the hadronic interaction models
around *10 T\e V }

\authors{M.~Amenomori$^{1}$, X.~J.~Bi$^{2}$, D.~Chen$^{3}$, W.~Y.~Chen$^{2}$, S.~W.~Cui$^{4}$,
Danzengluobu$^{5}$, L.~K.~Ding$^{2}$, X.~H.~Ding$^{5}$,
C.~F.~Feng$^{6}$, Zhaoyang Feng$^{2}$, Z.~Y.~Feng$^{7}$,
Q.~B.~Gou$^{2}$, H.~W.~Guo$^{5}$, Y.~Q.~Guo$^{2}$, H.~H.~He$^{2}$,
Z.~T.~He$^{4,2}$, K.~Hibino$^{8}$, N.~Hotta$^{9}$, Haibing~Hu$^{5}$,
H.~B.~Hu$^{2}$, J.~Huang$^{2}$, W.~J.~Li$^{2,7}$, H.~Y.~Jia$^{7}$,
L.~Jiang$^{2}$, F.~Kajino$^{10}$, K.~Kasahara$^{11}$,
Y.~Katayose$^{12}$, C.~Kato$^{13}$, K.~Kawata$^{3}$,
Labaciren$^{5}$, G.~M.~Le$^{2}$, A.~F.~Li$^{14,6,2}$, C.~Liu$^{2}$,
J.~S.~Liu$^{2}$, H.~Lu$^{2}$, X.~R.~Meng$^{5}$,
K.~Mizutani$^{11,15}$, K.~Munakata$^{13}$, H.~Nanjo$^{1}$,
M.~Nishizawa$^{16}$, M.~Ohnishi$^{3}$, I.~Ohta$^{17}$,
S.~Ozawa$^{11}$, X.~L.~Qian$^{6,2}$, X.~B.~Qu$^{2}$,
T.~Saito$^{18}$, T.~Y.~Saito$^{19}$, M.~Sakata$^{10}$,
T.~K.~Sako$^{12}$, J.~Shao$^{2,6}$, M.~Shibata$^{12}$,
A.~Shiomi$^{20}$, T.~Shirai$^{8}$, H.~Sugimoto$^{21}$,
M.~Takita$^{3}$, Y.~H.~Tan$^{2}$, N.~Tateyama$^{8}$,
S.~Torii$^{11}$, H.~Tsuchiya$^{22}$, S.~Udo$^{8}$, H.~Wang$^{2}$,
H.~R.~Wu$^{2}$, L.~Xue$^{6}$, Y.~Yamamoto$^{10}$, Z.~Yang$^{2}$,
S.~Yasue$^{23}$, A.~F.~Yuan$^{5}$, T.~Yuda$^{3}$, L.~M.~Zhai$^{2}$,
H.~M.~Zhang$^{2}$, J.~L.~Zhang$^{2}$, X.~Y.~Zhang$^{6}$,
Y.~Zhang$^{2}$, Yi~Zhang$^{2}$, Ying~Zhang$^{2}$,
Zhaxisangzhu$^{5}$, X.~X.~Zhou$^{7}$\ (The Tibet AS$\gamma$
Collaboration)}

\afiliations{$^{1}$Department of Physics, Hirosaki University, Hirosaki 036-8561, Japan\\
$^{2}$Key Laboratory of Particle Astrophysics, Institute of High
Energy Physics, Chinese
Academy of Sciences, Beijing 100049, China\\
$^{3}$Institute for Cosmic Ray Research, University of Tokyo, Kashiwa 277-8582, Japan\\
$^{4}$Department of Physics, Hebei Normal University, Shijiazhuang 050016, China\\
$^{5}$Department of Mathematics and Physics, Tibet University, Lhasa 850000, China\\
$^{6}$Department of Physics, Shandong University, Jinan 250100, China\\
$^{7}$Institute of Modern Physics, SouthWest Jiaotong University, Chengdu 610031, China\\
$^{8}$Faculty of Engineering, Kanagawa University, Yokohama 221-8686, Japan\\
$^{9}$Faculty of Education, Utsunomiya University, Utsunomiya 321-8505, Japan\\
$^{10}$Department of Physics, Konan University, Kobe 658-8501, Japan\\
$^{11}$Research Institute for Science and Engineering, Waseda
University, Tokyo 169-8555,
Japan\\
$^{12}$Faculty of Engineering, Yokohama National University, Yokohama 240-8501, Japan\\
$^{13}$Department of Physics, Shinshu University, Matsumoto 390-8621, Japan\\
$^{14}$School of Information Science and Engineering, Shandong
Agriculture University,
Taian 271018, China\\
$^{15}$Saitama University, Saitama 338-8570, Japan\\
$^{16}$National Institute of Informatics, Tokyo 101-8430, Japan\\
$^{17}$Sakushin Gakuin University, Utsunomiya 321-3295, Japan\\
$^{18}$Tokyo Metropolitan College of Industrial Technology, Tokyo 116-8523, Japan\\
$^{19}$Max-Planck-Institut f\"ur Physik, M\"unchen D-80805, Deutschland\\
$^{20}$College of Industrial Technology, Nihon University, Narashino 275-8576, Japan\\
$^{21}$Shonan Institute of Technology, Fujisawa 251-8511, Japan\\
$^{22}$RIKEN, Wako 351-0198, Japan\\
$^{23}$School of General Education, Shinshu University, Matsumoto
390-8621, Japan}

\abstract{A hybrid experiment has been started by AS$\gamma$
collaboration at Tibet, China, since May 2009, that consists of a
burst-detector-grid (YAC, Yangbajing Air shower Core array) and the
Tibet air-shower array (Tibet-III). The first step of YAC, called
¡°YAC-I¡±, contains 16 detector units and observes high energy
electromagnetic particles in air-shower cores within several meters
from the shower axis, and Tibet-III array measures the total energy
and the arrival direction of air showers. YAC-I  is used to check
hadronic interaction models currently used for air-shower
simulations such as QGSJET, SIBYLL , EPOS etc. through the
multi-parameter measurement in air-shower cores. In this paper, we
used a data set collected from May 1st 2009 through February 23rd
2010 by the YAC-I. The effective live time used for the present
analysis is 169.65 days. The preliminary results of the interaction
model checking at *10 TeV energy region is reported in this paper. }

\begin{document}
\maketitle

\section{Introduction}
\vspace{-0.3cm} The interpretation of the extensive air showers
(EAS) is known to inevitably depend on the Monte Carlo simulations
which are based on some hadronic interaction models and cosmic ray
composition models. At present, the simulation code CORSIKA that is
comprehensively used in the surface cosmic ray studies includes many
interaction models. For multi-parameter measurements of EASs, it is
known that no interaction model can explain all data consistently.
Therefore, the hadronic interaction models need to be checked and
further improved.

In this paper, we report our approach and the preliminary results to
check the hadronic interaction models QGSJET2 and SIBYLL2.1 at an
energy region of *10 TeV using the data obtained by the newly
constructed YAC-I (Yangbajing Air shower Core detector, the first
stage). The energy region of *10 TeV is chosen by the following
considerations: 1) The primary composition at this energies has been
better measured by direct measurements, and the uncertainty[1] is
smaller. 2) The corresponding energies in the center-of-mass system
are around *100 GeV, for that $Sp\bar{p}S$ collider made good
measurements on the inelastic cross section and on the particle
production in the near-forward region. The uncertainty from the
extrapolation to the very forward region is relatively small. 3) The
check of interaction models should be step-by-step executed from
lower energies to higher energies. 4) Due to the high altitude of
our Tibet experiment we could have good EAS core events recorded by
YAC-I at *10 TeV region.
\section{YAC-I experiment}
\vspace{-0.3cm}
 YAC-I, consisting of 16 scintillation detectors, has been
successfully operated at Yangbajing in Tibet, China, since May 2009,
 together with the Tibet-III EAS array. Each unit of YAC-I is
composed by a lead layer of 3.5 cm thickness ($\sim$7 r.l.) and a
plastic scintillator of size 40 cm $\times$50 cm$\times$1 cm. Each
YAC-I detector is used to record the electromagnetic showers induced
by high energy electrons and/or photons in the EAS cores. 16
detectors are arranged by $4\times4$, forming a $\sim$ 10 m$^2$
covering. To achieve a wider burst size measurement under the lead
layer, a wide dynamic range from 1 MIP (Minimum Ionization Particle)
to $10^6$ MIPs is demanded which is realized by two
photomultipliers(PMT),
 i.e. a high-gain PMT and a low-gain PMT for the range of
$1\sim3\times10^3$ and $10^3\sim10^6$ MIPs, respectively. The
response linearity of each YAC-I detector was calibrated by
cosmic-ray single muons and by the accelerator beam
(BEPC-LINAC)[10]. YAC-I is triggered when any one of 16 detectors
records a local shower with the size of at least 20 MIPs. The event
rate is about 30 Hz. The total live time of our data set in present
analysis is 169.65 days.
\begin{table*}[t]
\begin{center}
\begin{tabular}{l|cccccc}
\hline
Composition model    & Component & $1-10$ TeV  & $10-100 $ TeV  & $10^3$-$10^4$ TeV  \\
\hline
            & P         & 38.6\%        & 32.0\%          & 24.2\%                          \\
HD        & He        & 24.7\%        & 22.4\%         & 19.1\%                        \\
model     & M         & 24.7\%        & 27.1\%          & 27.3\%                       \\
            & Fe        & 10.4\%        & 18.5\%          & 29.4\%                       \\
\hline
              & P   & 47.3\%   & 31.1\%  & 26.3\%  \\
NLA         & He  & 30.2\%   & 25.1\%  & 28.9\%   \\
model       & M   & 30.3\%   & 32.4\%  & 34.4\%  \\
              & Fe  & 12.8\%   & 11.3\%  & 10.6\%  \\
\hline
\end{tabular}
\caption{The fractions of the components of the HD model
and the NLA model. }\label{table_single}
\end{center}
\end{table*}
\vspace{-0.3cm}

\section{Simulations}

 \begin{figure}[!t]
  \vspace{0.5mm}
  \centering
  \includegraphics[width=3.2 in]{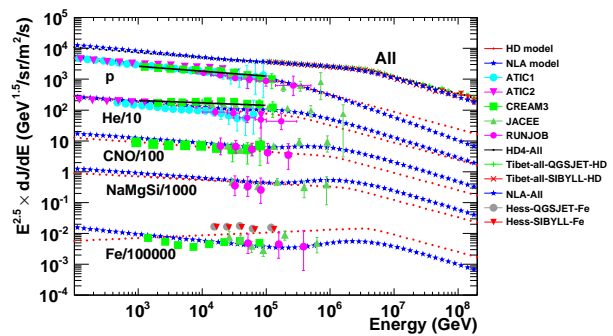}
  \caption{Primary cosmic-ray composition for the HD model and the NLA model. The all-particle spectrum, which is the
the sum of all components, is nomalized to the Tibet data and
compared with other experiments. Black solid lines denoted p and He
spectrum  that are fitted from PAMELA and CREAM data [6][8][9].
 }
  \label{double_fig}
 \end{figure}

\vspace{-0.2cm} \vspace{-0.3cm} A Monte Carlo simulation has been
carried out on the development of EASs in the atmosphere and their
response in YAC-I. The simulation code CORSIKA (version 6.204)[2]
including QGSJET2 and SIBYLL2.1 hadronic interaction models are used
to generate air shower events. In this work, two primary composition
models are used, the heavy dominant model(HD)[3] and the non-linear
acceleration model(NLA)[4], as shown in Fig.1. The proton spectrum
of the two models is connected with the direct experiment in the low
energy and consistent with the spectrum obtained from the Tibet
AS+EC experiment in the high energy. The He spectrum of HD model
coincides with the results from RUNJOB and ATIC-I, but the He
spectrum of NLA coincides with the results from JACEE, ATIC-II,
CREAM3. The sum of all single-component spectra can reproduce the
sharp knee in all particle spectrum[4]. The fractions of the
component of the two composition models in different energy regions
are listed in Table 1.

In the simulation, the primary energy is sampled from 1 TeV to
infinite with zenith angles from 0 to 60 degrees incident
isotropically. The axis of each EAS event is randomly dropped onto
an area of 32.84 m $\times $32.14 m with YAC-I at its central part.
When high energy electrons or photons hit a YAC-I detector, the
EPICS(8.64)[5] code is used to generate the cascade showers in the
Pb layer and in the detector. To identify an AS core event we use
following quantities:

$N_b$: number of shower particles recorded by a YAC-I detector;

$N_{hit}$: number of YAC-I detectors with $N_b$ higher than a
threshold value $N_b$$_{min}$ (If $N_b$$\geq$$N_b$$_{min}$ for a YAC
detector, it is called that this detector is fired);

$\sum$$N_b$: the sum of all $N_b$ from 1 to $N_{hit}$;

$N_b^{top}$:  the maximum of all $N_b$;

$<R>$: the mean lateral distance from the center of a fired detector
to the $N_b$ weighter center of all fired detectors;

$<N_b{\times}R>$: the mean $N_b$ weighted lateral distance from the
center of a fired detector to the $N_b$ weighted center.

By choosing a $N_b$$_{min}$ value and setting some event selection
conditions we can obtain different event samples for that primary
energies range at different region. For the present work using
$N_b$$_{min}$=200 and the conditions of 1) $N_{hit}$=3,
$\sum$$N_b\geq3500$, 2) $N_{hit}$ =4, $\sum$$N_b\geq1700$, 3)
$N_{hit}$=5, $\sum$$N_b\geq2200$, three Monte Carlo samples with the
mode energy at $\sim$35 TeV, $\sim$70 TeV and $\sim$90 TeV are
obtained, respectively. Their sample sizes are seen from Table 2. In
all 12 cases, Monte Carlo shows that the core resolution is better
than 2 m if the $N_b$ weighted center is taking as the AS core.
\begin{table*}[t]
\begin{center}
\begin{tabular}{l|cccccc}
\hline \small {Mode energy}     & $\sim35$ TeV & $\sim70$ TeV
 & $\sim90$ TeV  \\ \hline
\small{QGSJET2+HD}   & 2640 & 8893  & 5352  \\
\small{SIBYLL2.1+HD}  & 2921  & 9239  & 5495  \\
\small{QGSJET2+NLA}  & 4484  & 15515  & 9312 \\
\small{SIBYLL2.1+NLA} & 3941 & 12568  & 7492 \\
\small{YAC-I data} & 1773 & 5640 & 3220  \\
\hline
\end{tabular}
\caption{The numbers of core events of three samples selected by
conditions 1), 2), 3) (see Section 3)in Monte Carlo simulation and
YAC-I data. }\label{table_double}
\end{center}
\end{table*}

\section{Data analysis}

 \begin{figure}[!t]
  \vspace{1mm}
  \centering
  \includegraphics[width=2.6 in]{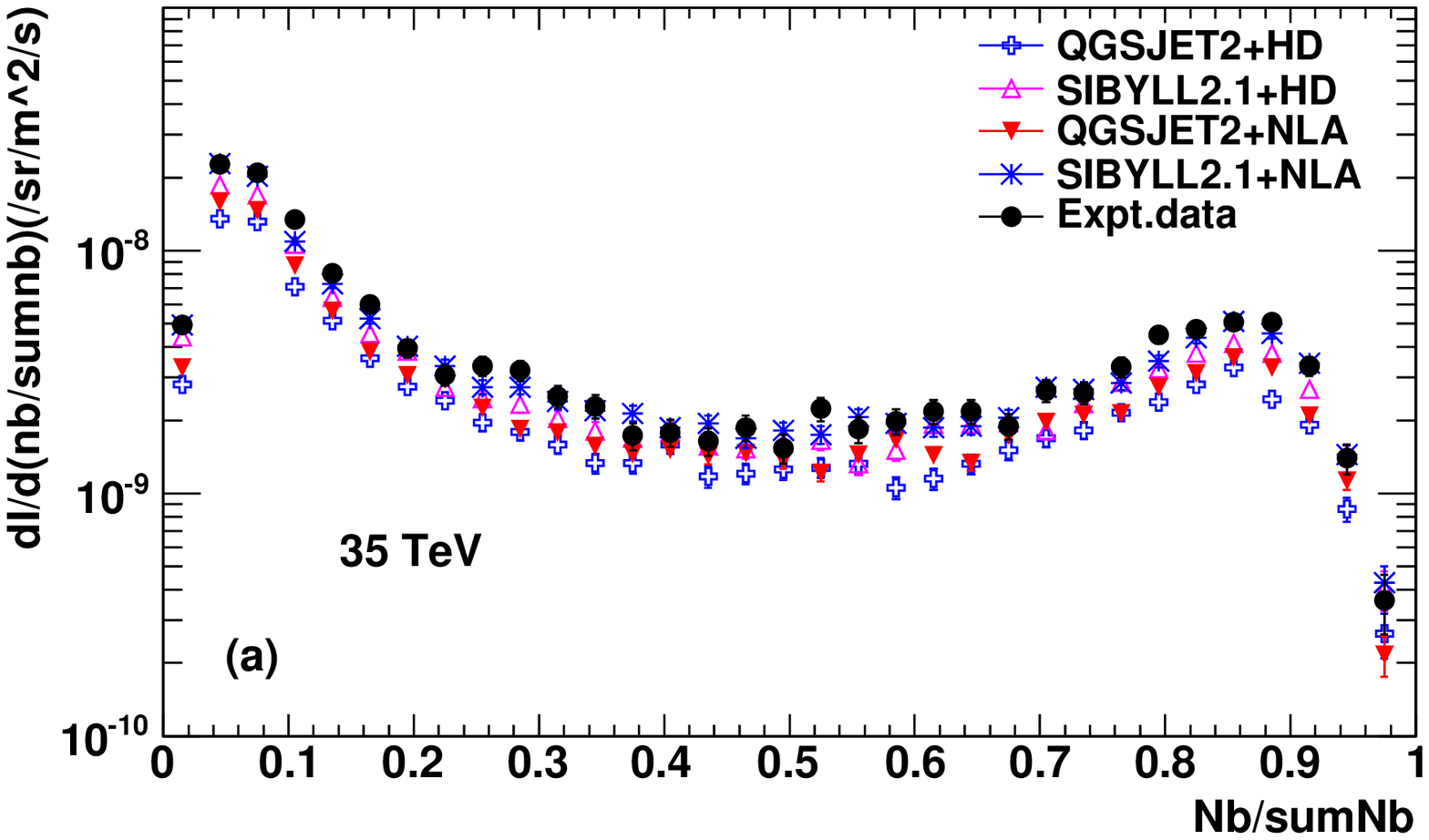}
  \includegraphics[width=2.6 in]{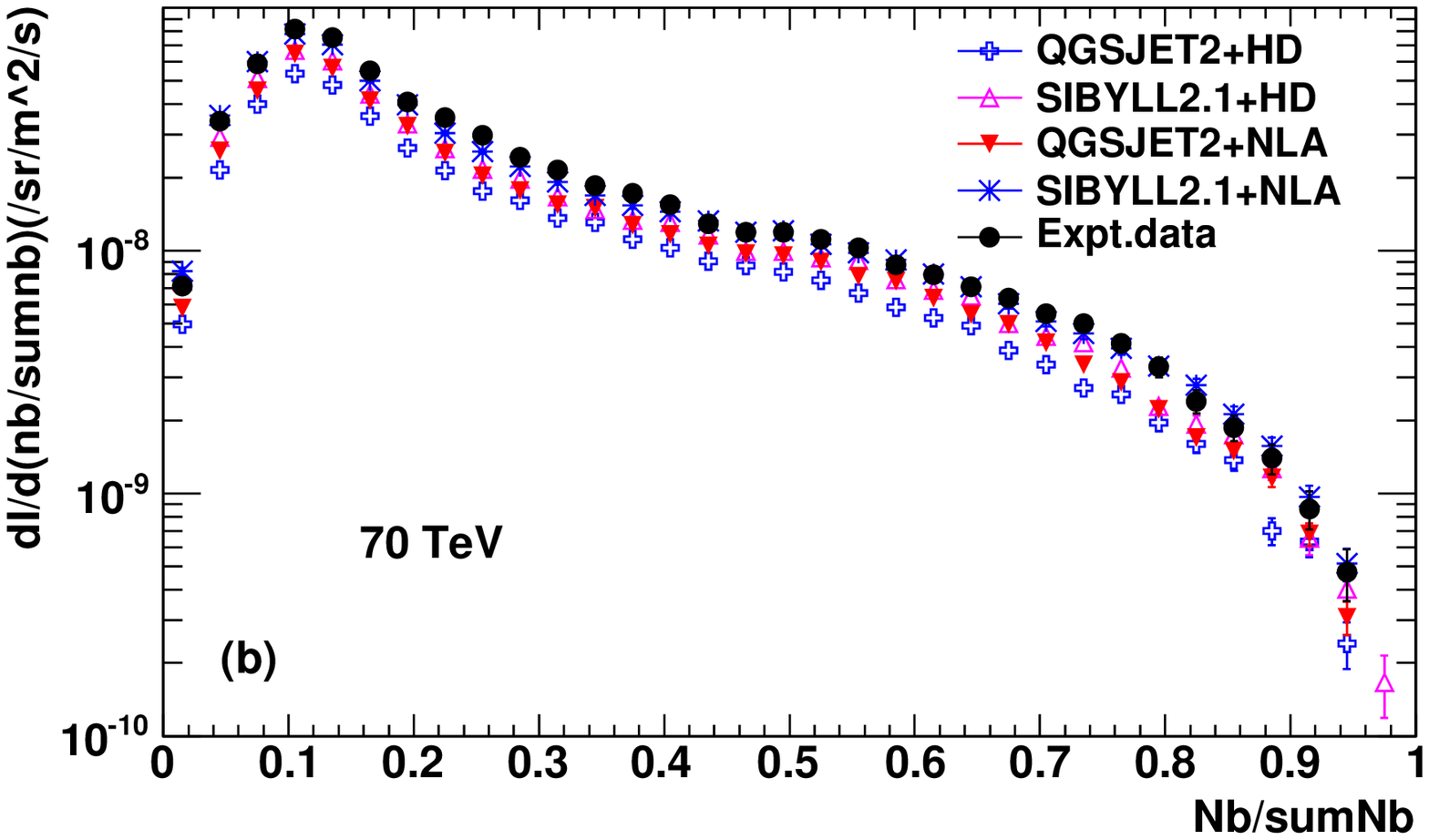}
  \includegraphics[width=2.6 in]{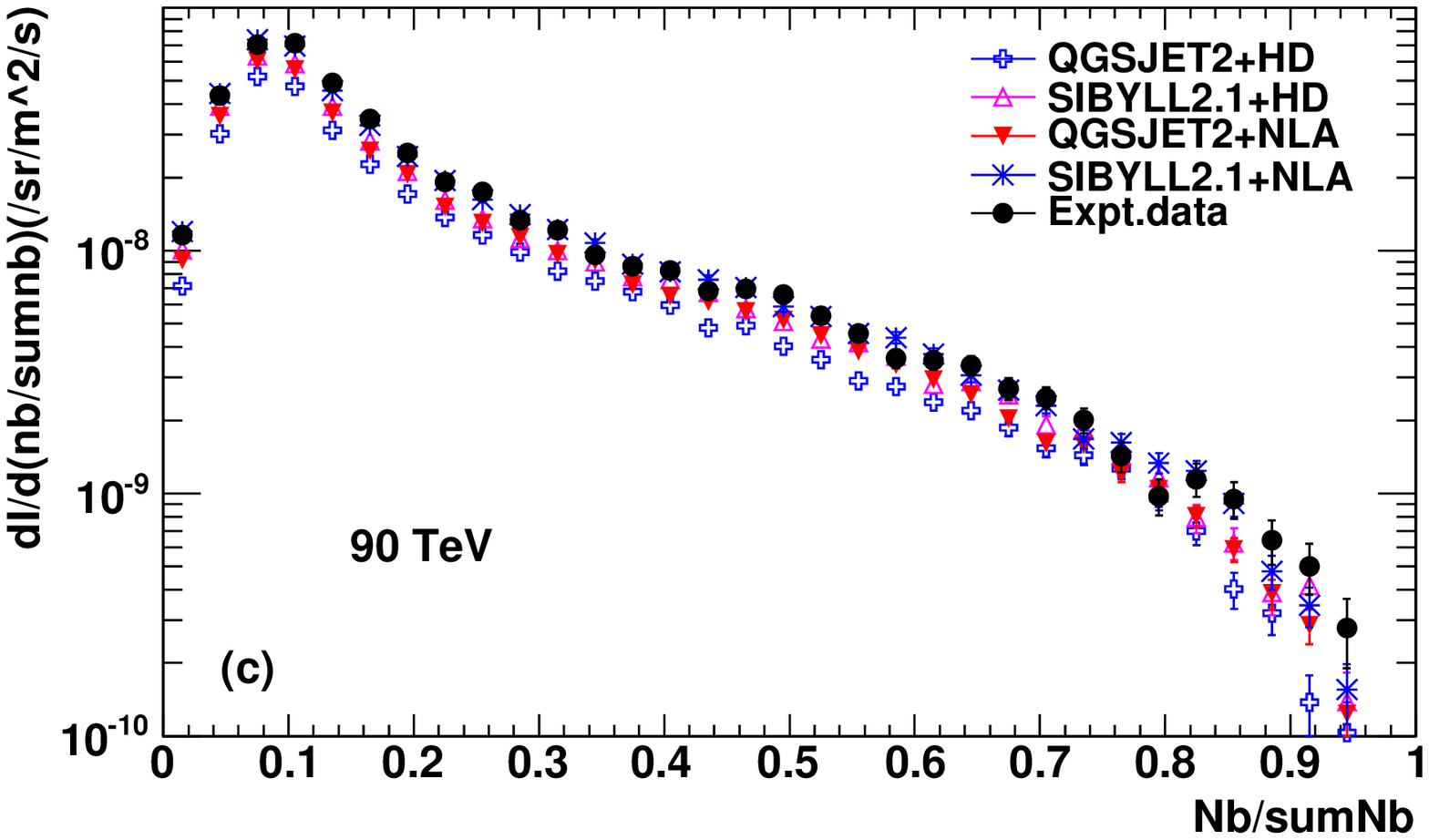}
  \caption{The absolute intensity of $N_b$/$\sum${$N_b$}
  distribution of the three samples for QGSJET2+HD model, SIBYLL2.1+HD model, QGSJET2+NLA model, SIBYLL2.1+NLA model and the
experimental data. }
  \label{double_fig}
 \end{figure}

 \begin{figure}[!t]
  \vspace{1mm}
  \centering
  \includegraphics[width=2.6 in]{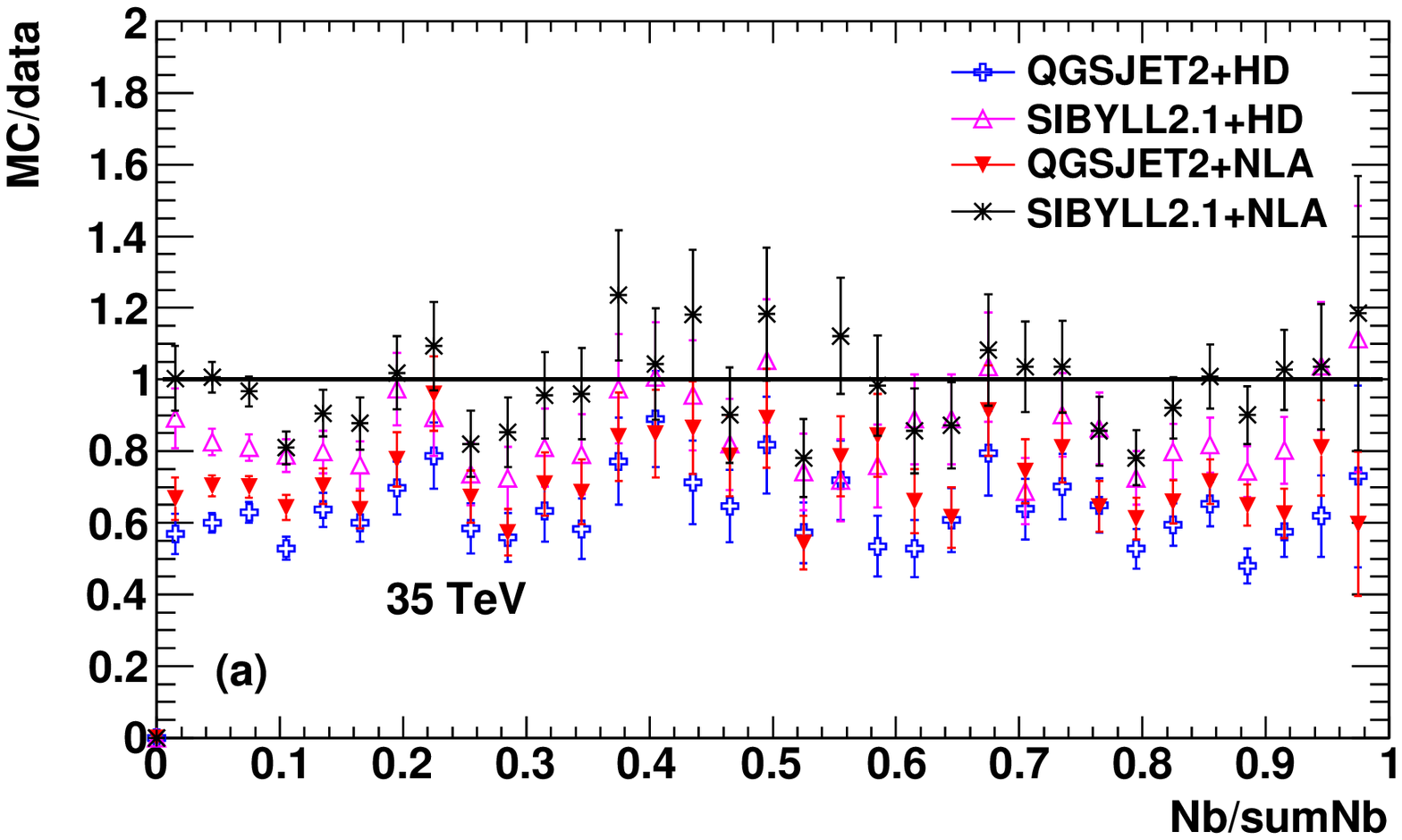}
  \includegraphics[width=2.6 in]{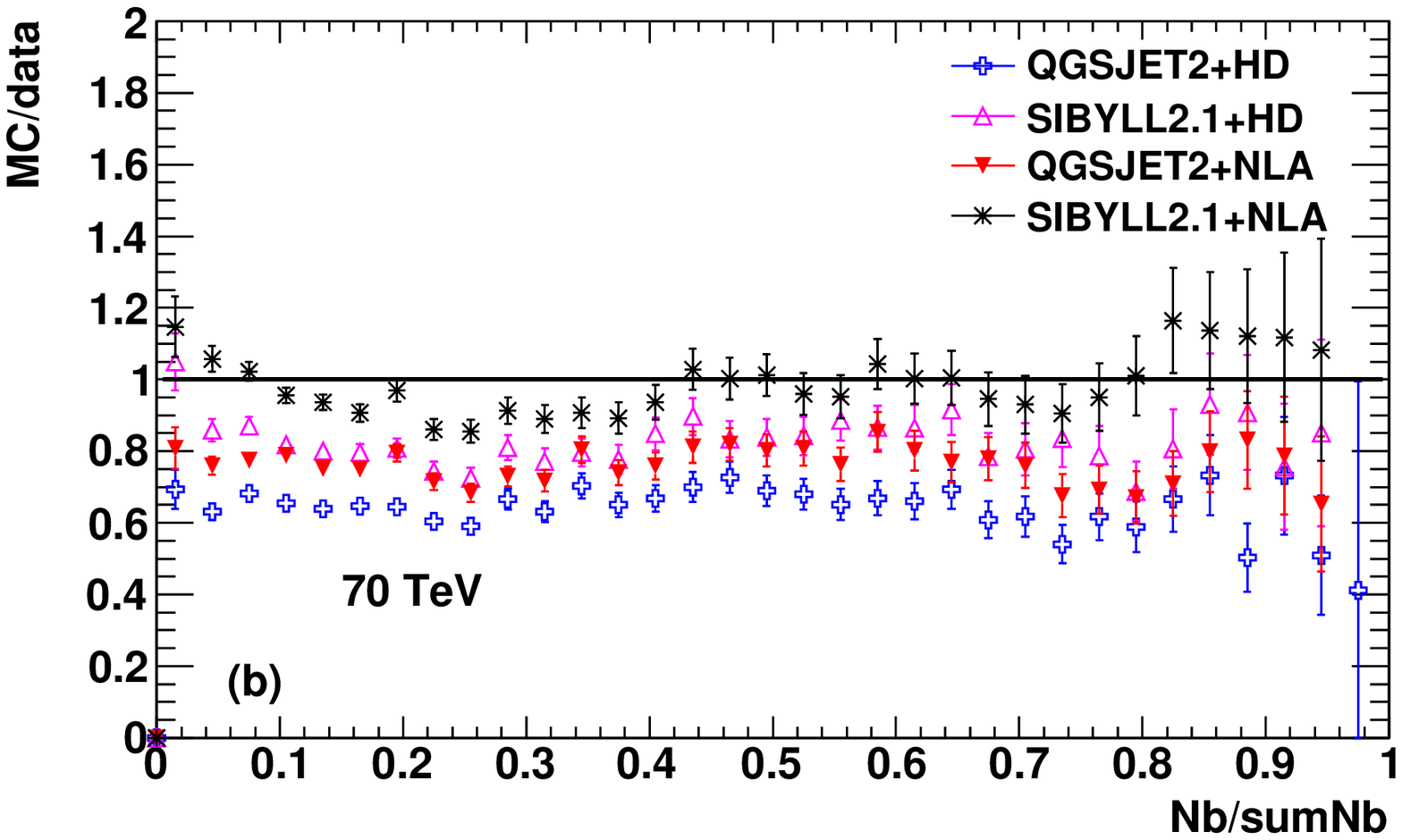}
  \includegraphics[width=2.6 in]{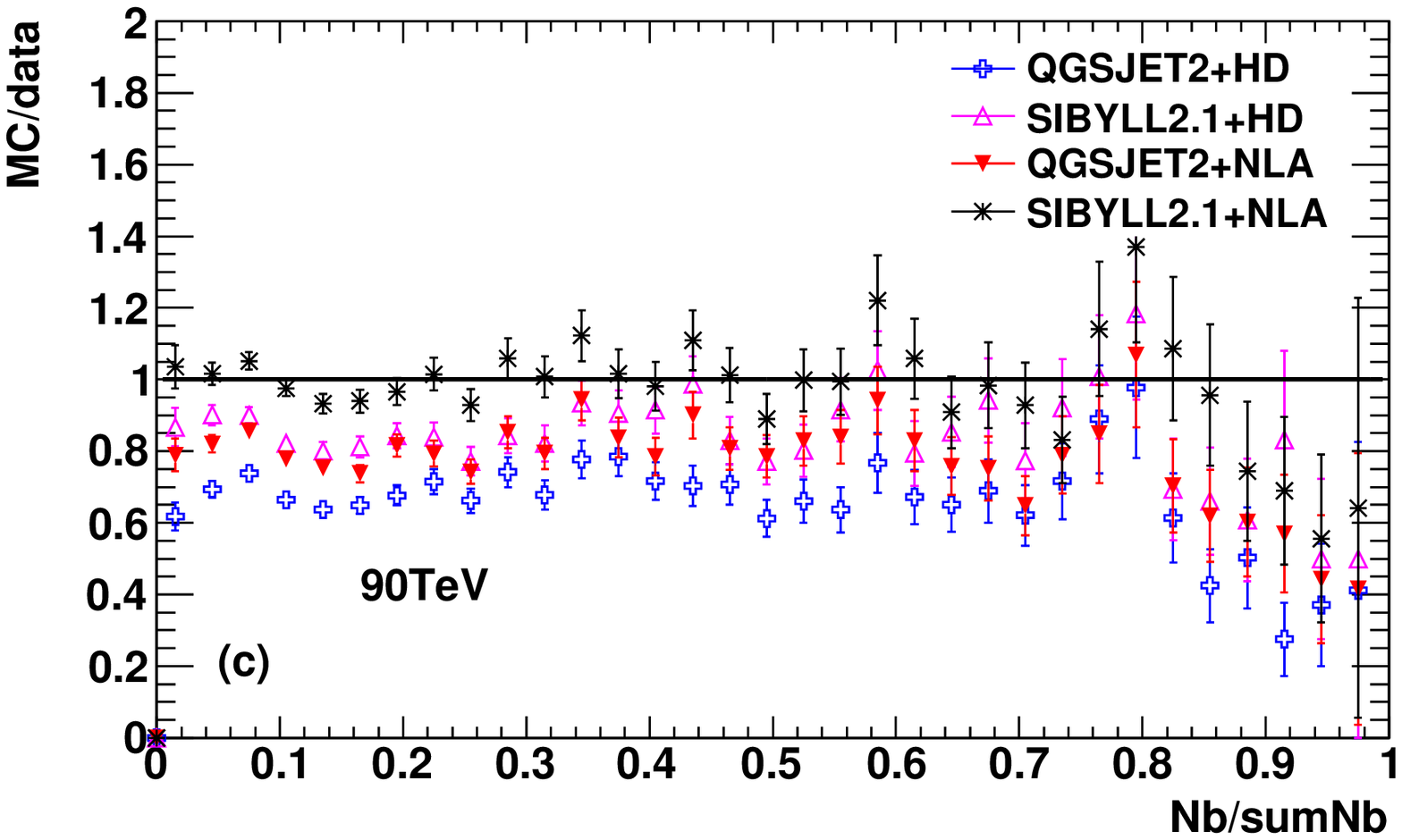}
  \caption{The ratio of absolute intensity
   between  MC and Exp. data. }
  \label{double_fig}
 \end{figure}
\vspace{-0.1cm}

\vspace{-0.2cm}
 The present analysis uses a subset of data collected
in the period from May 1st, 2009 to February 23rd, 2010, having
stable instrument conditions. We first check some noises appeared in
our data set due to the hardware and the environment conditions. It
is found that most noises appear as smaller 'signals'. Taking
$N_b$$_{min}$=200 can remove them and achieve a biggest available
data sample. In addition, we found some gain drift for some YAC-I
detectors during the operating process, by checking $N_b$ spectrum
of each YAC-I detector. By an 'off-line self-calibration' method
this effect is carefully treated and corrected. Then the fit between
high-gain signals and low-gain signals in their overlapping region
is executed and $N_b$ (or MIPs) is obtained for each fired YAC-I
detector.

After the off-line calibration, $\sim150000$ events with $N_b$
$\geq$ 200 and $N_{hit}$ $\geq$ 1 are obtained. Three experimental
data samples with the same selection criteria of Monte Carlo are
obtained. The sizes of each sample are also listed in Table2.
\section{Results and Discussion}
Since our Monte Carlo simulation is started from 1 TeV, in order to
normalize MC data and experimental data, we need to know the
integral intensity of all particles of cosmic rays at E$_{0}$ = 1
TeV. Starting from H$\ddot{o}$randal's spectra of each
composition[7], we improve the major 8 ones (p, He, C, O, Ne, Mg, Si
, and Fe) by the newest measurements [8][9][10]. The resultant
integral intensity: I(E$_{0}$$\geq$1 TeV) = 0.139
cm$^{-2}$s$^{-1}$sr$^{-1}$ with the error +0.0013, -0.0012 coming
from  the error of the index of each of the 8 spectra.

The comparison of our data with Monte Carlo simulations of
QGSJET2+NLA, QGSJET2+HD, SIBYLL2.1+NLA and SIBYLL2.1+HD is seen from
Fig.2 and Fig.3:

 (1) The shape of the distributions of $N_b$/$\sum$$N_b$ are consistent between
 the YAC-I data and simulation data in all four cases, indicating that
in the *10 TeV energy region the particle production spectrum of
QGSJET2 and SIBYLL2.1 may correctly reflect the reality within our
experimental systematic uncertainty of a level about 10\%.

 (2) The comparison of event absolute intensities in all cases, as
seen from Fig.2 and Fig.3, shows some
discrepancies. The smallest one is SIBYLL2.1+NLA and the most
obvious one is QGSJET2+HD. For a given interaction model the NLA
composition is better than HD. For a given composition  model
SIBYLL2.1 is better than QGSJET2.

 (3) But note that, as seen from Fig 1, both composition
models NLA and HD used a steeper He spectrum, if comparing with the
new results from PAMELA and CREAM. It can be estimated that NLA
under-estimated the number of He events for about 30-40\% in the
related energy region. If involving this factor, QGSJET2 results can
go higher and be possible better consistent with data. A further
analysis is going on.
\section{Summary}
\vspace{-0.1cm}

A smaller high-altitude AS core detector YAC-I shows the ability and
sensitivity in checking the hadronic interaction models in
*10 TeV region.

The experimental distributions, $N_b$/$\sum$$N_b$ and has the shape
very close to the Monte Carlo predictions of QGSJET2+NLA,
QGSJET2+HD, SIBYLL2.1+NLA and SIBYLL2.1+HD.  Some other quantities,
such as $N_b$, $\sum$$N_b$, $N_b$$^{top}$, Rw, Rw$\times$$N_b$ have
the same behavior as well, though we did not show them in this paper
due to the limit of the space.

Some discrepancies in the absolute intensities are seen. Data
normally shows a higher intensity than Monte Carlo. Taking a more
hard He spectrum (and somewhat hard proton spectrum) as given by
CREAM at the 1-100 TeV region can improve this situation. A further
study is going on.
\section{Acknowledgement}
\vspace{-0.2cm} This work is supported by the Chinese Academy of
Sciences (H9291450S3) and  the  Key  Laboratory of  Particle
Astrophysics, Institute of High Energy Physics, CAS.  The Knowledge
Innovation Fund (H95451D0U2 and H8515530U1) of IHEP, China. The
project Y0293900TF of NSFC also provide support to this study.


\clearpage


\begin{thebibliography}{}
\bibitem{1} B. Wiebel-Sooth, et al., Astronomy and Astrophysics 330, 389
            (1998).
\bibitem{2} D.Heck et al., Report FZKA 6019,(1998);\\
            J.Knapp, D.Heck et al., Report FZKA 3640,(1997).
\bibitem{3} M.Amenomori, et al., ApJ 678:1165-1179,(2008).
\bibitem{4} M.Shibata, J.Huang et al., ApJ 716, 1076-1083,(2010).
\bibitem{5} K.Kasahara, et al., http://eweb.b6.kanagawa-u.ac.jp \\
            /$\sim$Kasahara/ResearchHome/EPICSHome/Index.html.
\bibitem{6} H.S.Ahn et.al., ApJL, 714:l89-l93,(2010).
\bibitem{7} J.R. H$\ddot{o}$randel, Astrop. Phys., 19,192,(2003).
\bibitem{8} O.Adriani, et al., Science 332,69,(2011).
\bibitem{9} H.S.Ahn, et al., ApJ., 707,593-603,(2009).
\bibitem{10} M.Amenomori, et al., ICRC32(HE 1.4,ID: 1217), (2011).





\end{thebibliography}
\end{document}